\documentclass[twocolumn,showpacs,preprintnumbers,amsmath,amssymb,prb]{revtex4}
\usepackage{graphicx}% Include figure files
\usepackage{dcolumn}% Align table columns on decimal point
\usepackage{bm}% bold math
\usepackage{tabularx}% Table width
\usepackage{amsmath}% Higher math
\usepackage{float}

\begin{document}

\title{Specific heat and thermal conductivity of ferromagnetic magnons in Yttrium Iron Garnet}

\author{B. Y. Pan, T. Y. Guan, X. C. Hong, S. Y. Zhou, X. Qiu, H. Zhang, and S. Y. Li$^*$}

\affiliation{State Key Laboratory of Surface Physics, Department of
Physics, and Laboratory of Advanced Materials, Fudan University,
Shanghai 200433, P. R. China}

\date{\today}
\begin{abstract}
The specific heat and thermal conductivity of the insulating
ferrimagnet Y$_3$Fe$_5$O$_{12}$ (Yttrium Iron Garnet, YIG) single
crystal were measured down to 50 mK. The ferromagnetic magnon
specific heat $C$$_m$ shows a characteristic $T^{1.5}$ dependence
down to 0.77 K. Below 0.77 K, a downward deviation is observed, which is
attributed to the magnetic dipole-dipole interaction with typical
magnitude of 10$^{-4}$ eV.  The ferromagnetic magnon thermal
conductivity $\kappa_m$ does not show the characteristic $T^2$ dependence
below 0.8 K. To fit the $\kappa_m$ data, both magnetic defect
scattering effect and dipole-dipole interaction are taken into
account. These results complete our understanding of the thermodynamic
and thermal transport properties of the low-lying ferromagnetic
magnons.
\end{abstract}


\maketitle

\section{Introduction}
Recently, the ferrimagntic insulator Y$_3$Fe$_5$O$_{12}$ (Yttrium
Iron Garnet, YIG) draws great attention due to the long-range
transport ability of spin current.\cite{YK,HiK} In these
experiments, electronic signal can be transferred by spin angular
momentum of the spin waves in insulating YIG, via the spin Hall and
inverse spin Hall effects, which provides a new method to transfer
information by pure spin waves.\cite{YK,HiK} In this context, it is
important to understand the thermodynamic and transport properties
of the spin waves.

In spin wave theory for magnets, the quanta of spin waves are
magnons. There are antiferromagnetic (AFM) and ferromagnetic (FM)
magnons, which have totally different dispersion relations, and
distinct thermodynamic and transport properties. The AFM magnons
have linear dispersion relation, so both their specific heat and
boundary-limited thermal conductivity at low temperature obey the
$T^3$ dependence.\cite{UR,SYL} For FM magnons, the situation is more
complex due to the existence of magnetic dipole-dipole interaction
(MDDI).\cite{IO,SHC}

At not very low temperature, the dispersion relation $E$ = $Dk^2$ is
a good approximation, and the specific heat and boundary-limited
thermal conductivity of FM magnons show the characteristic $T^{1.5}$
and $T^2$ dependence, respectively.\cite{HS,AIA,AK} The equations
are
\begin{equation}
C_m(T)=\frac{15\zeta(5/2)k_B^{2.5}T^{1.5}}{32{\pi}^{1.5}D^{1.5}}
\end{equation}
and
\begin{equation}
\kappa_m(T)=\frac{\zeta(3)k_B^3LT^2}{{\pi}^2{\hbar}D},
\end{equation}
where $\zeta$ is the Riemann function, $L$ is the boundary-limited
mean free path.\cite{UR,HS,AIA,AK} However, at sub-Kelvin temperature range, the MDDI
\begin{equation*}
H_{d-d}=2\mu_B^2\sum_{i\neq{j}}\frac{r_{ij}^2(\bf{S_i}\bf{S_j})-(\bf{r_j}\bf{S_i})(\bf{r}_{ij}\bf{S}_j)}{r_{ij}^5},
\end{equation*}
with typical order of $10^{-4}$ eV, has to be considered for FM
magnons.\cite{VC} It significantly modifies the dispersion relation
of FM magnons below 1 K, and makes the approximate form $E$ = $Dk^2$
no longer valid.\cite{TH}

The MDDI is long range and anisotropic. It is a basic interaction in
magnets and critical for many phenomena such as the demagnetization
factors and the formation of domain walls in ferromagnets, the spin
ice behavior in Ising pyrochlore magnets, and the spin anisotropy in
ferromagntic films.\cite{NWA,BCDH,JGG} Theoretically, when the
effect of MDDI was taken into account for FM magnons, both the
$T^{1.5}$ dependence of $C_m$ and $T^2$ dependence of $\kappa_m$
changed.\cite{IO,SHC,VC,DCM} Yet so far experimental verification of
this effect on FM magnons is still lacking.

YIG is an archetypical ferrimagnetic insulator. Its low-energy
magnetic excitations are FM magnons because at low temperature the
important spin-wave branch in a ferrite has the same form as in a
ferromagnet.\cite{HK,DD} The study of this interesting and complex
compound started from late 1950s, and it has become indispensable
for investigating the properties of magnets since then.\cite{VC} In
fact, YIG is the first material in which thermal conductivity
contributed by  magnetic excitations was observed.\cite{BL} At low
temperature, its magnon specific heat and thermal conductivity can
even exceed those contributed by phonons.\cite{BL,RLD,DW} Therefore,
YIG is an ideal compound to study the thermodynamic and transport
properties of FM magnons.

Previous lowest temperature for specific heat measurement on YIG was
1 K.\cite{TDE,JEK} The data of magnon specific heat $C_m(T)$ between
1 and 4 K showed the characteristic $T^{1.5}$
dependence.\cite{TDE,JEK} The thermal conductivity of YIG was
roughly measured down to 0.23 K.\cite{DW} The data of magnon thermal
conductivity $\kappa_m(T)$ between 0.23 and 1 K did not obey the
characteristic $T^2$ dependence, and it was explained by considering
the effect of magnetic defect scattering.\cite{DW,JC,JC2}

In this paper, we present the specific heat and thermal conductivity
measurements of YIG single crystal down to 50 mK. The magnon
specific heat data deviate downward from the $T^{1.5}$ dependence
below 0.77 K, which is attributed to the MDDI effect. Below 0.2 K,
the magnon thermal conductivity data cannot be fitted by only
considering the boundary and magnetic defect scatterings, and the
MDDI effect has to be taken into account. To our knowlege, this is
the first experimental observation of MDDI effect on FM magnon
specific heat and thermal conductivity, giving complete
understanding of the thermodynamic and thermal transport properties
of the low-lying FM magnons.
\section{Experiment}
\begin{figure}
\includegraphics[clip,width=8.5cm]{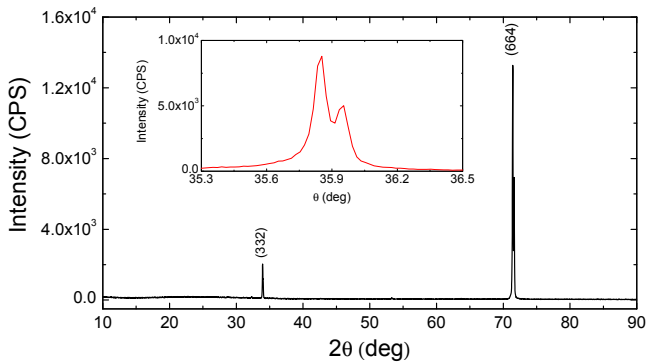}
\caption{(Color online) X-ray diffraction pattern for the (332)
plane of YIG single crystal. Inset: rocking curve of the (664)
reflection. The two peaks are from  the Cu $K_{\alpha 1}$ and
$K_{\alpha 2}$ radiations, respectively}
\end{figure}
The YIG single crystal was grown by the optical floating zone
furnace. \cite{LLA,SK,SK2,SK3,AR} The single crystal grew along the
[332] crystallographic direction, characterized by the X-ray
diffraction. Ultra-low temperature specific heat measurement was
carried out on a sample with mass 45.55 mg in a small dilution
refrigerator adapted into a Physical Property Measurement System
(PPMS, Quantum Design company).

Sample 1 (S1) was cut from the single crystal for thermal
conductivity measurements. It is rectangle shaped with dimensions
2.08$\times$0.84 mm$^2$ in the plane and 0.63 mm thick along the
[332] direction (the sample growth direction). Sample 2 (S2) was
obtained by thinning S1 to 0.23 mm. For both samples, the heat
current was along the [11$\overline{3}$] direction.

Ultra-low temperature thermal conductivity was measured in a
dilution refrigerator (Oxford Instruments), using a standard
four-wire steady-state method with two RuO$_2$ chip thermometers,
calibrated $in$ $situ$ against a reference RuO$_2$ thermometer. Four
contacts were made on the sample surface by silver epoxy. Magnetic
fields were applied parallel to the heat current.

\section{Results and Discussion}

The quality of our YIG single crystal was characterized by X-ray
diffraction (XRD), as shown in Fig. 1. The main panel is the XRD
pattern of the (332) plane. The inset shows the rocking scan curve
of the (664) reflection, with two peaks from the Cu $K_{\alpha 1}$
and $K_{\alpha 2}$ radiations, respectively. The full width at half
maximum (FWHM) of the peak from Cu $K_{\alpha 1}$ is only 0.07
$^{\rm{o}}$, indicating high quality of the crystal.

\begin{figure}
\includegraphics[clip,width=7.5cm]{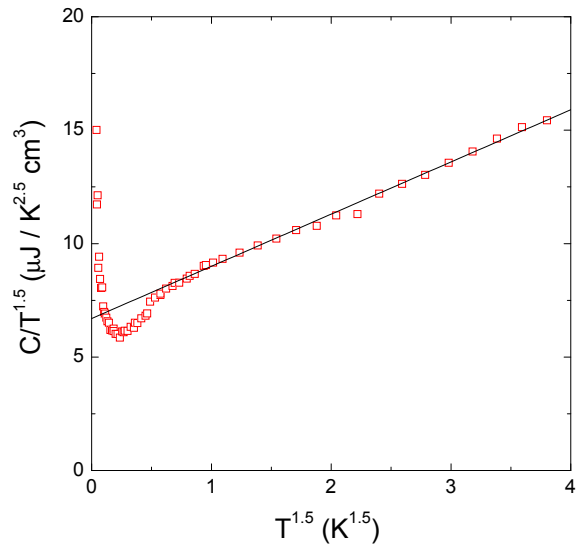}
\caption{(Color online) Specific heat of YIG single crystal. The
data between 0.77 and 2.5 K can be fitted by the solid line $C =
6.7T^{1.5}+2.3T^3$. Below 0.77 K the curve deviates from the solid
line, which is attributed to the effect of magnetic dipole-dipole
interaction. Note that the upturn below 0.38 K is the Schottky
anomaly from nuclear moments.}
\end{figure}

The specific heat of YIG single crystal below 2.5 K is shown in Fig.
2. In the figure, $C/T^{1.5}$ was poltted as a function of $T^{1.5}$
in order to sperate phonon and magnon contributions. Between 0.77
and 2.5 K, the total specific heat can be fitted by $C$ =
$aT^{1.5}$+$bT^3$, in which the first term is from the FM magnons
and the second term is from the phonons. From the fitted coefficient
$a$ = 6.7 $\mu$J/cm$^3$ K$^{2.5}$, we get $D$ =
5.2$\times$10$^{-36}$ J cm$^2$ according to Eq. (1). This value is
very close to Edmonds and Petersen's result $D$ =
5.1$\times$10$^{-36}$ J cm$^2$.\cite{TDE}

Below 0.77 K, however, there is an apparent deviation from the solid
fitting line. As we have described, the MDDI will affect the
specific heat of FM magnons below 1 K, by modifying the dispersion
relation. The modified dispersion relation was proposed as the
following form\cite{TH}
\begin{equation}
E(k)=\sqrt{(Dk^2-N_z\hbar\omega_m)(Dk^2-N_z\hbar\omega_m+\hbar\omega_m\sin^2\theta_k)}
\end{equation}
where $\theta_k$ is the angle between the magnon wave vector and the
magnetization direction, $N$$_z$ is the $z$ demagnetization factor,
and $\hbar$$\omega$$_m$ = $g$$\mu_B$4$\pi$$M$. The dispersion
relation of Eq. (3) is plotted in Fig. 3. The approximate dispersion
relation $E$ = $Dk^2$ is only valid for $k_B$$T$ $\gg$
$\hbar$$\omega$$_m$. For YIG, 4$\pi$$M$ = 2449 Gs,\cite{DTE,MAG} $g$
= 2,\cite{EPW} so $\hbar$$\omega$$_m$  = $g\mu_B$4$\pi$$M$ =
0.32$k_B$. Therefore, the $T^{1.5}$ dependence of $C_m$ should only
hold for $T$ $\gg$ 0.32 K. From our experimental data in Fig. 2,
$C_m$ shows $T^{1.5}$ dependence above 0.77 K. The downward
deviation below 0.77 K should come from the effect of MDDI. Note
that the upturn below 0.38 K is the Schottky anomaly from nuclear
moments.

\begin{figure}
\includegraphics[clip,width=9cm]{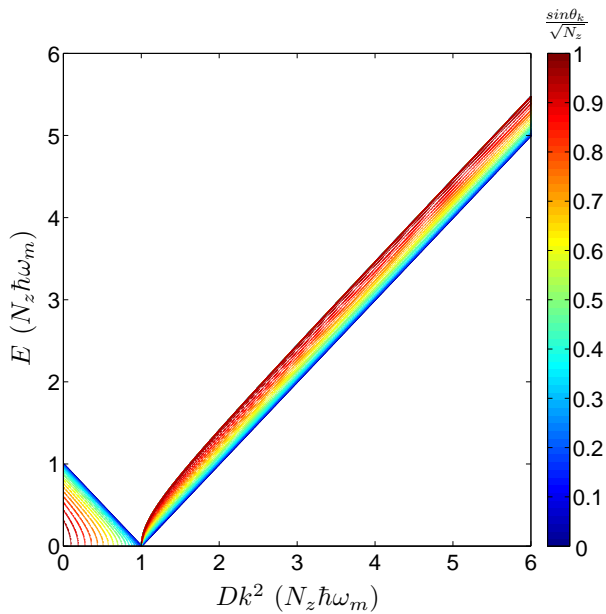}
\caption{(Color online) The dispersion relation of FM magnons at low
energy when considering the magnetic dipole-dipole interaction
effect:\cite{IO} $E(k)$ =
$\sqrt{(Dk^2-N_z\hbar\omega_m)(Dk^2-N_z\hbar\omega_m+\hbar\omega_m\sin^2\theta_k)}$.
It strongly deviates from the $E$ = $Dk^2$ approximation below the
energy $\hbar\omega_m$ = $g\mu_B$4$\pi$$M$. For YIG, $\hbar\omega_m$
= 0.32$k_B$.}
\end{figure}

Next we discuss the thermal conductivity results. Figure 4(a) shows
the ultra-low temperature thermal conductivity of YIG in magnetic
fields up to 11 T, plotted as $\kappa/T$ vs $T$. One can see that
$\kappa$ is strongly suppressed by field, as previously
reported.\cite{BL,RLD,DW} In the insulating YIG, the total thermal
conductivity can be expressed as $\kappa = \kappa_{ph} + \kappa_m$,
in which $\kappa_{ph}$ and $\kappa_m$ are the phonon and magnon
thermal conductivity, respectively. Since $\kappa_{ph}$ is usually
not affected by magnetic field, the rapid suppression of $\kappa$
with field in Fig. 4(a) should come from the reduction of
$\kappa_m$. As we know, the external magnetic field $H$ opens a gap
$\Delta$ = g$\mu_B$$H$ in the magnon spectrum. When external field
is high enough to satisfy g$\mu_B$$H$ $\gg$ $\kappa_B$$T$, there
would be no magnon contribution. The saturated thermal conductivity
from $H$ = 4 to 11 T shown in Fig. 4(a) suggests that there is only
phonon contribution left below 0.8 K in $H \geq$ 4 T, consistent
with previous reports.\cite{RLD,DW}

Therefore the FM magnon thermal conductivity $\kappa_m$ in zero
field can be extracted by subtracting $\kappa$(4T) from
$\kappa$(0T). In Fig. 4(b), $\kappa_m$ is plotted as $\kappa_m/T$ vs
$T$. For AFM magnons, ballistic boundary-limited $\kappa_m = aT^3$
was observed below 0.5 K.\cite{SYL} Apparently, for FM magnons in
Fig. 4(b), there is no such a simple power-law temperature
dependence.

\begin{figure}
\includegraphics[clip,width=7.5cm]{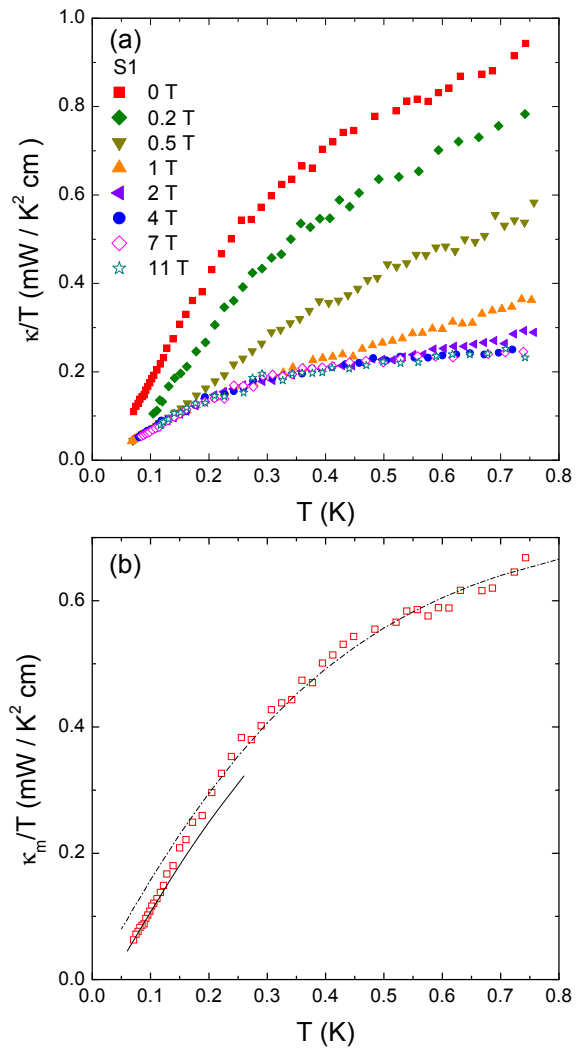}
\caption{(Color online) (a) Thermal conductivity of YIG single
crystal in magnetic fields up to 11 T. In $H \geq$ 4 T, $\kappa$
tends to saturate. (b) Zero field thermal conductivity of FM magnons
obtained by subtracting $\kappa$(4T) from $\kappa$(0T). The dashed
line is the fitting of the data between 0.2 and 0.8 K, by
considering the boundary scattering and magnetic defect scattering.
The deviation below 0.2 K is attributed to the MDDI effect. The
solid is the fitting of the data below 0.12 K by including the MDDI
effect.}
\end{figure}

\begin{figure}
\includegraphics[clip,width=7.5cm]{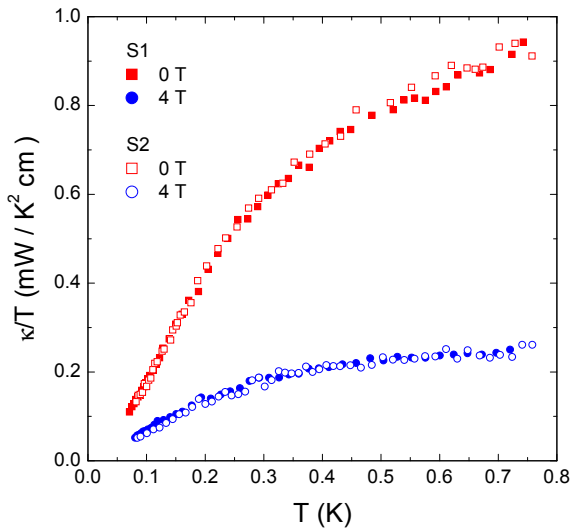}
\caption{(Color online) Thermal conductivity of YIG samples S1 and
S2 in zero field and $H$ = 4 T. S2 is obtained by thinning S1 from
0.63 to 0.23 mm.}
\end{figure}

Theoretically, the thermal conductivity of magnons is calculated by
the equation
\begin{equation}
\kappa=\frac{k_B}{24{\pi}^3\hbar}{\int}(\frac{E}{k_BT})^2\frac{e^{E/k_BT}}{(e^{E/k_BT}-1)^2}(\nabla_kE){\ell}d^3k,
\end{equation}
where $\ell$ is the mean free path of magnons. At not very low
temperature, if the approximate dispersion relation $E$ = $Dk^2$ of
FM magnons is taken and only boundary scattering is considered, we
get the $T^2$ dependence of $\kappa_m$ in Eq. (2). Therefore, the
anomalous temperature dependence of $\kappa_m$ in Fig. 4(b) must
come from the MDDI effect or some other scattering mechanism.
Previously, additional magnetic defect scattering was considered,
and the data of $\kappa_m$ between 0.23 and 1 K can be well
fitted.\cite{DW} In this case, $\ell^{-1}$ can be expressed as the
sum of two terms
\begin{equation}
\ell^{-1}=L^{-1}+\ell_D^{-1}
\end{equation}
where $L^{-1}$ is from boundary scattering and $\ell_D^{-1}$ is from
magnetic defect scattering with $\ell_D^{-1}$ = $\alpha$$k^4$. In
this way, the magnon thermal conductivity is
\begin{equation}
\kappa_m(T)=BT^2\int_{0}^{\infty}\\\frac{x^3csch^2(\frac{1}{2}x)dx}{1+{\beta}T^2x^2},
\end{equation}
where $x=E/k_BT$, $B = \frac{\zeta(3)k_B^3L}{{\pi}^2{\hbar}D}$,
$\beta = \frac{{\alpha}Lk^2_B}{D^2}$. By using this model, our
$\kappa_m$ data between 0.2 and 0.8 K can also be well fitted, with
the parameters $B$ = 0.056 mW/K$^3$ cm and $\beta$ = 0.15 K$^{-2}$.
Using the obtained value of $B$, together with $D$ =
5.2$\times$10$^{-36}$ J cm$^2$, we get the boundary-limited mean
free path $L$ = 32.7 $\mu$m, which we will discuss later.

However, the magnetic defect scattering model can not fit the
$\kappa_m$ data below 0.2 K, as seen in Fig. 4(b). We further
consider the effect of MDDI. Since the dispersion relation Eq. (3)
is too complex to do calculation, Ortenberger and Sparks chose a
simplified dispersion relation\cite{IO}
\begin{equation}
E=Dk^2+ck_B.
\end{equation}
Substituting this dispersion relation
into Eq. (4) results in
\begin{equation}
\kappa_m=BT\int_{0}^{\infty}\frac{x^2csch^2(\frac{1}{2}x)(Tx+c)}{1+\beta(Tx+c)^2}dx.
\end{equation}
With the $B$ and $\beta$ value obtained above, the $\kappa_m$ data
can be fitted well below 0.12 K, as shown in Fig. 4(b). We want to
emphasize that we tried to fit the $\kappa_m$ data below 0.8 K with
only boundary scattering and MDDI effect, without considering the
magnetic defect scattering, but it did not work. Therefore, the
magnetic defect scattering mechanism is necessary for explain the
low-temperature $\kappa_m$ data in YIG. Comparing the results of
$C_m$ and $\kappa_m$, the MDDI starts to affect thermal conductivity
at a lower temperature than specific heat. The reason may relate to
the involvement of magnetic defect scattering in thermal
conductivity.

Finally, we discuss the boundary-limited mean free path $L$ of FM
magnons in YIG. From Fig. 4(b), $L$ = 32.7 $\mu$m is obtained by the
fitting. This value is one order smaller than the expected $L$ =
2$\sqrt{A/\pi}$ = 727 $\mu$m for sample S1, with $A$ being the cross
section area.\cite{SYL2} Such a phenomenon has been observed
previously.\cite{BL,RLD,SAF} Friedberg and Harris speculated that it
is due to the inner boundary of thin layers rich in Fe$^{\rm{2+}}$
inside the sample.\cite{SAF} To test this idea, we did the same
thermal conductivity measurements on sample S2, obtained by thinning
S1 from 0.63 to 0.23 mm. In Fig. 5, the 0 and 4 T data of S2 are
almost identical to those of S1, indicating the same $\kappa_m$ in
S1 and S2. This result shows that $\kappa_m$ indeed does not change
with the sample boundary, therefore the actual boundaries are inside
the sample, likely the thin layers proposed by Friedberg and
Harris.\cite{SAF} These thin layers may form during the growing
process.

\section{Summary}

In summary, by extending the measurements of the specific heat and
thermal conductivity down to 50 mK, we investigate the thermodynamic
and transport properties of the low-lying ferromagnetic magnons in
YIG single crystal. The deviation of magnon specific heat $C_m(T)$
from the characteristic $T^{1.5}$ dependence below 0.77 K is
attributed to the effect of magnetic dipole-dipole interaction. The
magnon thermal conductivity $\kappa_m(T)$ is extracted by
subtracting $\kappa_m$(4T) from $\kappa_m$(0T). Below 0.8 K,
$\kappa_m(T)$ does not obey the characteristic $T^2$ dependence due
to the magnetic defect scattering. With further decreasing
temperature, the magnetic dipole-dipole interaction also shows
effect on $\kappa_m(T)$ below 0.2 K. Our work provides a complete
understanding of the thermodynamic and transport properties of the
low-lying ferromagnetic magnons.

\begin{center}
{\bf ACKNOWLEDGEMENTS}
\end{center}

This work is supported by the Natural Science Foundation of China,
the Ministry of Science and Technology of China (National Basic
Research Program No: 2009CB929203 and 2012CB821402), Program for
Professor of Special Appointment (Eastern Scholar) at Shanghai
Institutions of Higher Learning.\\

$^*$ E-mail: shiyan$\_$li@fudan.edu.cn

\end{document}